\documentclass[aps,prl,floatfix,showpacs,twocolumn,superscriptaddress,tighten]{revtex4}

\usepackage{amsmath}
\usepackage{graphicx}
\usepackage{color}

\newcommand{\ket}[1]{\mbox{$|#1\rangle$}}
\newcommand{\bra}[1]{\mbox{$\langle #1|$}}

\newcommand{\ketbra}[2]{\mbox{$|#1\rangle\langle #2|$}}
\newcommand{\op}[1]{\mbox{\boldmath $\hat{#1}$}}

\definecolor{back}{named}{Apricot}

\newcommand{\mo}[3][]{{\op{#2}_{#3}^{\dagger #1}}}

\newcommand{\vac}{\ket{\mathbf{0}}}

\begin{document}

\title{Input states for quantum gates}

\author{A.~Gilchrist}
\email{alexei@physics.uq.edu.au}
\affiliation{Center for Quantum Computer Technology and Department of 
Physics, University of Queensland, QLD 4072, Brisbane, Australia.}

\author{W.J.~Munro}
\affiliation{Hewlett-Packard Laboratories, Filton Road, Stoke 
Gifford, Bristol, BS34 8QZ, United Kingdom.}

\author{A.G.~White}
\affiliation{Center for Quantum Computer Technology and Department of 
Physics, University of Queensland, QLD 4072, Brisbane, Australia.}

\date{\today}

\begin{abstract}
  We examine three possible implementations of non-deterministic
  linear optical \textsc{cnot} gates with a view to an in-principle
  demonstration in the near future. To this end we consider
  demonstrating the gates using currently available sources such as
  spontaneous parametric down conversion and coherent states, and
  current detectors only able to distinguish between zero or many
  photons.  The demonstration is possible in the co-incidence basis
  and the errors introduced by the non-optimal input states and
  detectors are analysed.
\end{abstract}

\pacs{03.67.Lx,42.50.-p}

\maketitle

\section{Introduction}

Optics is a natural candidate for implementing a variety of quantum
information protocols. Photons make beguiling qubits: at optical
frequencies the qubits are largely decoupled from the environment and so
experience little decoherence, and single qubit gates are easily realised
via passive optical elements. Some protocols, notably quantum computation,
also require two-qubit gates. Until recently this was regarded as optically
infeasible, since the required nonlinear interaction is much greater than
that available with extant materials. However, it is now widely recognised
that the necessary nonlinearity can be realised non-deterministically via
measurement, and that deterministic gates can be achieved by combining such
non-deterministic gates and teleportation \cite{01klm46}.

There are a number of proposals for implementing a non-deterministic
\textsc{cnot} gate with linear optics and photodetectors
\cite{01klm46,02ralph012314,0111092,ndCNOT2,ndCNOT3,0107091}. The
proposals require deterministic, or heralded, single photon sources,
and/or \emph{selective} detectors, that can distinguish with very high
efficiency between zero, one and multiple photons. Current commercial
optical sources and detectors fall well short of these capabilities.
Although there are a number of active research programs aimed at
producing both efficient selective detectors
\cite{02imamoglu163602,02james183601}, and deterministic photon
sources \cite{qDOT1,qDOT2,diamond}, nonselective avalanche
photodiodes, spontaneous parametric downconversion (SPDC) and coherent
states remain the best accessible laboratory options.  While we could
side-step the single photon source problem by using an SPDC source
conditioned on the detection of a photon in one arm if we had
selective detectors, demonstrating a four-photon \textsc{cnot} gate
without quantum memory would be frustratingly slow.

In this paper we examine three proposals which allow a \textsc{cnot}
to be implemented non-destructively on the control and target modes,
to ascertain under what conditions it is possible to
demonstrate and characterise the gates operation using SPDC sources,
coherent states and \emph{non-selective} detectors (detectors only
able to resolve zero and multiple photons). The aim is to identify a scheme
that allows a scalable \textsc{cnot} implementation to be initially examined 
with current sources and detectors, and into which we can easily incorporate
single photon sources and selective detectors as they become available.

Typically the gates involve four photons with the qubit states are
encoded in the polarisation state of the control and target modes $c$
and $t$, and the \textsc{cnot} operation is implemented with the aid
of some ancillary modes $a$, $b$ etc. We will consider starting with
the control and target modes each in a general superposition (we could
also consider initially entangled states though these may be more
difficult experimentally)
\begin{equation}
   \label{eq:init-fock}
\ket{\psi_\mathrm{in}}_{ct}=(A_h\mo{c}{h}+A_v\mo{c}{v})(B_h\mo{t}{h}+B_v\mo{t}{v})\vac
\end{equation}
with $|A_h|^2+|A_v|^2=|B_h|^2+|B_v|^2=1$, and where $\mo{c}{h,v}$ and
$\mo{t}{h,v}$ are bosonic creation operators for mode $c_{h,v}$ and
$t_{h,v}$ etc. In the interest of brevity we will use the notation above
where we write the state in terms of creation operators acting on the
vacuum state.

The modes are first entangled with a linear optics network
$U_\textsc{cnot}$ comprised of beamsplitters , phase shifters,
waveplates, and polarising beam splitters. Finally the gate is
conditioned on detecting the ancillary modes in some appropriate
state, which leaves the state of the control and target modes as if a
\textsc{cnot} had been applied.

The key simplification for our purposes is to detect in the
`coincidence basis' --- where we detect the output of the ancillary
modes and also of the target and control modes and postselect out
those events that do not simultaneously register a photon in all four
modes.  The advantage of this configuration is that now we can use
\emph{non-selective} detectors, since if we get a ``click'' on all
four detectors we've accounted for all the photons in the system. This
is a much less stringent requirement on the detectors and in
particular can be fulfilled by existing avalanche photodiodes.  We
model the non-selective detectors with a positive-operator-valued
measure (POVM), with the POVM elements associated with detecting
no photons or photons (one or more) simply being $\Pi_0=\ketbra{0}{0}$ 
and $\Pi_m= \sum_{n=1}^\infty \ketbra{n}{n}$ respectively.

The output state of a type-I SPDC can be described as
\begin{eqnarray}
\ket{\lambda}&=&\mathcal{M}_\lambda(\ket{00}+\lambda\ket{11}+\lambda^2\ket{22}+\cdots)\\
&=&\mathcal{M}_\lambda
\sum_{\small\begin{array}{c}n=0\\ \mathrm{(even)}\end{array}}^\infty
\frac{(\lambda\mo{a}{}\mo{b}{})^{\frac{n}{2}}}{\frac{n}{2}!}\vac
\end{eqnarray}
where $\mathcal{M}_\lambda=(1-\lambda^2)^{-1}$ and the sum is over
even $n$ where $n$ is the number of photons in each term. 

Now suppose that our input state to the optical circuit is some initial
pure state $\ket{\psi_\mathrm{in}}$, and that after passing through the
linear optical elements we are left in the state
$\ket{\psi_\mathrm{out}}=U_\textsc{cnot}\ket{\psi_\mathrm{in}}$.  The
probability that we get a count simultaneously in modes $c$, $t$, $a$ and
$b$ with non-selective detectors is
\begin{equation}
   P = \bra{\psi_\mathrm{out}}\Pi_m^{(c)}\otimes\Pi_m^{(t)}\otimes
\Pi_m^{(a)}\otimes\Pi_m^{(b)}\ket{\psi_\mathrm{out}}
\end{equation}
For the ideal case where we had single photon inputs to the gate, we
will label this probability as $P_1$. We can now introduce the
``single photon visibility'' as a figure of merit for how close the
gate operates to the ideal:
\begin{equation}
  \label{eq:visibility}
\mathcal{V} = \frac{1}{2}\left(\frac{s-e}{s+e}+1\right)
\end{equation}
where $s$ is the product of the probability of obtaining the single
photon terms from the source, with $P_1$ the probability of the gate
functioning.  The ``error'' $e=\max (s-P)$ where $P$ is the actual
probability of obtaining a count on the detectors. The maximisation is
over all qubit input states to the gate. Hence if the error totally
dominates the visibility is close to zero, if the noise is small the
visibility is close to one. As a guide a visibility of $0.8$ corresponds
to an error a quarter of the size of the single photon ``signal'' $s$.

\section{Simplified KLM CNOT}

In the originally proposed non-deterministic \textsc{cnot} gate
\cite{01klm46} the nonlinear sign shift elements were interferometric:
these elements can be replaced by sequential beamsplitters to make a
simplified \textsc{cnot} gate \cite{02ralph012314}, one example of
which is
\begin{eqnarray}
   \label{eq:sKLM}
\op{U}_{\textsc{sklm}}&=&\op{B}_{t_ht_v}(\frac{\pi}{4})\op{B}_{c_vt_h}(\frac{\pi}{4})
\op{B}_{bt_h}(\theta_2)\op{B}_{ac_v}(\theta_2)\nonumber\\
&&{}\op{B}_{c_vt_h}(\frac{\pi}{4})\op{B}_{t_hv_2}(\theta_1)\op{B}_{t_ht_v}(\frac{\pi}{4})
\op{B}_{v_1c_v}(\theta_1)
\end{eqnarray}
where $ \op{B}_{ab}$ represents a beam splitter with the following action
\begin{eqnarray}
   \op{B}_{ab}(\theta)\op{a}\op{B}^\dagger_{ab}(\theta) &=& 
\op{a}\cos\theta+\op{b}\sin\theta\\
   \op{B}_{ab}(\theta)\op{b}\op{B}^\dagger_{ab}(\theta) &=& 
\op{a}\sin\theta -\op{b}\cos\theta
\end{eqnarray}
and $\cos^2\theta$ is the reflectivity. The angle choices for the gate
are given by $\theta_1=\cos^{-1}\sqrt{5-3\sqrt{2}}$ and
$\theta_2=\cos^{-1}\sqrt{(3-\sqrt{2})/7}$; $c$ and $t$ are the control
and target modes and $a$, $b$, $v_1$ and $v_2$ are independent ancillary
modes.  The gate is conditioned on detecting a single photon in the
modes $a$ and $b$ and detecting no photons in the modes $v_1$ and
$v_2$.


Consider the case where both the control, target and ancillary photons
are supplied by two independent SPDC sources.  The input state is
$\ket{\lambda}_{ct}\ket{\epsilon}_{ab}$ which  can be written as a sum
over total photon number
\begin{eqnarray}
\label{eq:2SPDC}
  \ket{\phi_\mathrm{in}}&=& \mathcal{M}_\lambda\mathcal{M}_\epsilon \sum_{\small
\begin{array}{c}n=0\\\mathrm{(even)}\end{array}}^\infty\op{Q}_n\vac\\
\op{Q}_n&=& \sum_{m=0}^{\frac{n}{2}}
\frac{\epsilon^m\lambda^{\frac{n}{2}-m}}{m!(\frac{n}{2}-m)!}(\mo{a}{}\mo{b}{})^m(\mo{c}{}\mo{t}{})^{\frac{n}{2}-m}\nonumber
\end{eqnarray}
The control and target horizontal and vertical polarisation modes are
then each mixed on a beamsplitter so that we achieve the input state
\eqref{eq:init-fock} for those modes.

Since we are
postselecting on getting a `click' at four detectors then the terms with
$n<4$ will always get postselected out. Similarly, the terms with $n>4$
will get postselected out if we used selective detectors otherwise they
represent error terms. In the latter case, so long as 
$\epsilon,\lambda\ll1$ these terms will be
small. For the case were $n=4$, three input terms contribute:
\begin{equation}
\label{eq:in4}
\ket{\psi_\mathrm{in}^{(4)}}=(\lambda\epsilon\mo{a}{}\mo{b}{}\mo{c}{}\mo{t}{}
+\frac{\lambda^2}{2!}\mo[2]{c}{}\mo[2]{t}{}+\frac{\epsilon^2}{2!}\mo[2]{a}{}\mo[2]{b}{})\vac
\end{equation}
While the first of these terms is equivalent to having four initial
Fock states, the remaining two terms have the possibility of surviving
the postselection criteria and skewing the statistics observed.
Fortunately these last two terms lead to output terms which \emph{all}
get postselected out in the coincidence basis (e.g. two photons in the
control mode).  This means that with selective detectors we could in
principle postselect out all terms that do not correspond to single
photon inputs from the output statistics. With non-selective detectors
the error terms will scale at least as $\lambda^3$ in amplitude (due
to the $n>4$ terms) so the figure of merit will scale with $\lambda$
(taking $\epsilon=\lambda$) as $\mathcal{V}\sim1/(1+\lambda^2)$ and
$\lambda$ is typically very small.

Now consider the situation where a SPDC supplies the two photons for
the control and target modes and weak coherent states are used for the
ancillary modes. The input state is then
$\ket{\phi_\mathrm{in}}=\ket{\lambda,\alpha,\beta}$ where $\mo{a}{}$
and $\mo{b}{}$ will be the creation operators for the coherent states.
After rearranging the state as a primary sum over photon number we get
\begin{equation}
  \ket{\phi_\mathrm{in}}=\sum_{n=0}^\infty
\sum_{\small\begin{array}{c}p=0\\ \mathrm{(even)}\end{array}}^n
\sum_{q=0}^{n-p}
\frac{(\lambda\mo{c}{}\mo{t}{})^{\frac{p}{2}}}{\frac{p}{2}!}
\frac{(\alpha\mo{a}{})^q}{q!}
\frac{(\beta\mo{b}{})^{n-p-q}}{(n-p-q)!}\vac
\end{equation}

Again, terms with $n<4$ will get postselected out and terms with $n>4$
will be weak error terms. The extra freedom from two independent
coherent states means that now there will be nine terms with $n=4$ and
only one of these is equivalent to using single photon inputs.

The terms were a single coherent state supplies all the photons always
gets postselected out. By setting $\beta=i\alpha$ the two terms where
a single coherent state supplies two photons and the paramp supplies
two will cancel each other due to the symmetry in the circuit. Finally
the term where the paramp supplies all the photons is postselected out
as before. This means that we will still get errors arising from the
input terms:
\begin{equation}
   \frac{i\alpha^4}{6}(\mo[3]{a}{}\mo{b}{} - \mo{a}{}\mo[3]{b}{})\vac
\end{equation}
Note that these do not depend on the input state that is encoded on
the control and target modes and by setting $\alpha\ll\lambda$ we can
scale away these terms relative the single photon terms. Unfortunately
this means that we cannot beat the photon collection rate that could
be achieved using two independent SPDC sources.

It should be noted that all the observations made for the simplified
KLM \textsc{cnot} also hold for the full KLM \textsc{cnot} in the
coincidence basis. However from the perspective of an initial
demonstration of the gate the simplified version is more desirable.
In the following two sections we will compare these results against
two other implementations of optical \textsc{cnot} gates.

\section{Entangled ancilla CNOT}

\begin{figure}
  \begin{center}
  \includegraphics[width=.7\columnwidth]{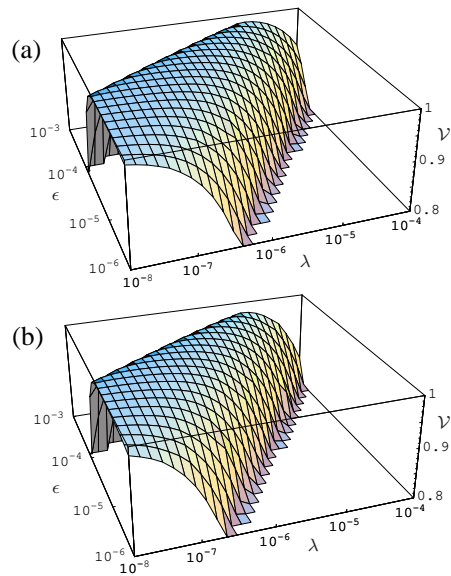}
  \end{center}
  \caption{The single photon visibility with non-selective detectors 
    as a function of the strengths of the SPDC sources. (a) the
    entangled ancilla gate, (b) the Knill gate. In both cases the
    input state was truncated at six photon terms, and the
    maximisation of the error was performed numerically.}
  \label{fig:combined}
\end{figure}

In a recent paper, Pittman, Jacobs and Franson \cite{0107091} proposed
using entangled ancilla to further simplify implementation of the
\textsc{cnot}. Consider that we have at our disposal an entangled state
$\ket{\phi}=(\op{a}_h\op{b}_h+\op{a}_v\op{b}_v)/\sqrt{2}\vac$, then we can
implemented the \textsc{cnot} between modes $c$ and $t$
by first applying the unitary 
\begin{equation}
  \op{U}_{\mathrm{ent}} =
\op{P}_{bd} \op{P}_{ae} \op{W}_a \op{W}_t \op{W}_b \op{P}_{bt} \op{W}_t \op{W}_b \op{P}_{ac}
\end{equation}
where $\op{W}_a$ represents a half-wave plate on mode $a$ and
$\op{P}_{ab}$ is a polarising beam splitter in modes $a$ and $b$ with
the effect that $a_h\rightarrow a_h$, $b_h\rightarrow b_h$,
$a_v\rightarrow b_v$, and $b_v\rightarrow a_v$. Finally the resulting
state is then conditioned on detecting a single photon in modes $a$
and $b$.  The raw success probability of this gate is $1/16$ which
rises to $1/4$ if fast feed-forward and correction is used.

Consider that the entangled pair in modes $a$ and $b$ are provided by
two type-I parametric downconverting crystals sandwiched together.
We'll fix the relative phase to get a particular Bell pair for the two
photon term:
\begin{eqnarray}
\ket{\epsilon_2}&=&\mathcal{M}_\epsilon^2(\ket{00}+\epsilon\ket{11}+\cdots)(\ket{00}+\epsilon\ket{11}+\cdots)\nonumber\\
&=&\mathcal{M}_\epsilon^2[\cdots+\epsilon(\ket{0011}+\ket{1100})+\cdots]
\end{eqnarray}
where the modes are $a_h$, $b_h$, $a_v$, and $b_v$ respectively.  Such
sources have been previously built and provide a relatively bright
source of polarisation entangled photons \cite{99kwwae773,99wjek3103}.
We can write this source succinctly as
\begin{eqnarray}
\ket{\epsilon_2}&=&\mathcal{M}_\epsilon^2\sum_{\small
    \begin{array}{c}n=0\\ \mathrm{(even)}\end{array}}^\infty \op{L}_n\vac\\
\op{L}_n&=&\sum_{m=0}^{n/2}\frac{\epsilon^\frac{n}{2}(\mo{a}{h}\mo{b}{h})^m
(\mo{a}{v}\mo{b}{v})^{\frac{n}{2}-m}}{m!(\frac{n}{2}-m)!}
\end{eqnarray}

With another independent paramp, $\ket{\lambda}$, supplying the
photons for the control and target modes, the input state becomes
\begin{equation}
\ket{\phi_\mathrm{in}}\equiv\mathcal{M}_\epsilon^2 \mathcal{M}_\lambda
\sum_{\small\begin{array}{c}n=0\\ \mathrm{(even)}\end{array}}^\infty
\sum_{\small\begin{array}{c}q=0\\ \mathrm{(even)}\end{array}}^n \op{L}_q
\frac{\lambda^\frac{n-q}{2}(\mo{c}{}\mo{t}{})^{\frac{n-q}{2}}}
{\frac{n-q}{2}!}
\end{equation}
where we will encode the qubits in the polarisation state
of the control and target modes, as in \eqref{eq:init-fock}.

Again all terms with $n<4$ will get postselected out. There are
six terms with $n=4$ of which two terms represents our single
photon input terms, the rest are error terms due to the sources.
With non-selective detectors terms with $n>4$ will also
contribute to the error.


The four photon terms in the output state that do not get postselected
out are
\begin{eqnarray}
\ket{out}&=&\frac{1}{2\sqrt{2}}\lambda \mo{a}{}\mo{b}{}(
A_vB_h \epsilon \mo{c}{v}\mo{t}{v} + A_vB_v \epsilon\mo{c}{v}\mo{t}{h}\nonumber\\
&&{}+A_h [A_vB_h^2\lambda-A_vB_v^2\lambda+B_v\epsilon]\mo{c}{h}\mo{t}{v}\nonumber\\
&&{}+A_h [A_vB_h^2\lambda-A_vB_v^2\lambda+B_h\epsilon]\mo{c}{h}\mo{t}{h}
)\vac
\end{eqnarray}
and by making $\lambda\ll\epsilon$ we can recover the single photon
terms and the action of the \textsc{cnot} with selective detectors.
This of course means that the count rate with this gate would be
considerably less than with the simplified KLM gate.  With
non-selective detectors, if we make $\lambda$ too small the error due
to the six photon input terms will dominate, so there is an optimum
$\lambda$ for a given $\epsilon$ see figure~\ref{fig:combined} (a).

There does not appear to be a way of using two coherent states to
replace one of the SPDC sources. If we replace either the control or
target mode then it is hard to see how the $\ket{02}$ and $\ket{20}$
terms could cancel as with the simplified KLM \textsc{cnot} since
these terms will have factors that depend on the encoded qubit.
Similarly replacing the source of entangled photons would then mean we
would have to entangle the single photon components which is difficult.
 
\section{Knill CNOT}

A recent numerical search for optical gates by Knill yielded a
\textsc{cnot} gate \cite{0110144} which operates with a probability of
$2/27$ and is
described by the following unitary,
\begin{eqnarray}
\op{U}_{\mathrm{Knill}} &=&
\op{B}_{t_vt_h}(\frac{\pi}{4})\op{B}_{ab}(\theta_3)\op{B}_{c_vt_v}(\theta_2)
\op{B}_{t_vb}(\theta_1)\nonumber\\
&&{}\op{B}_{c_va}(\theta_1)\op{B}_{t_vt_h}(\frac{\pi}{4})\op{F}_a(\pi)
\end{eqnarray}
where $\op{F}_a(\theta)$ is a phaseshift of $\theta$ on mode $a$ and
the reflectivities are given by $\theta_1=\cos^{-1}\sqrt{1/3}$,
$\theta_2=-\theta_1$ and $\theta_3=\cos^{-1}\sqrt{1/2+1/\sqrt{6}}$.
The gate requires two ancillary modes $a$ and $b$ initially in Fock
states to be finally detected also in single Fock states.


Consider the case where both the control, target and ancillary photons
are supplied by two independent SPDC sources.  The input state is
given by \eqref{eq:2SPDC} with the usual qubit encoding as in
equation~\eqref{eq:init-fock}.  We will again get the three terms
\eqref{eq:in4} possibly contributing to the error for $n=4$.  The last
term again leads to output terms which all get postselected out in the
coincidence basis.  Unfortunately the output terms produced by the
second term do not get postselected out leading to inherent errors in
the statistics we will observe. Notice however that all these terms
will be proportional to $\lambda^2$ so again by making
$\lambda\ll\epsilon$ we can scale these terms away with selective
detectors at the expense of the count rate. With non-selective
detectors there will again be an optimum $\lambda$, see
figure~\ref{fig:combined} (b), which is very similar to the previous gate.

\section{Conclusion}

We have examined three possible implementations for linear optics
\textsc{cnot} gates with a view to experimentally demonstrating their
operation in the near future. In considering demonstrating the gates
with SPDC and coherent state sources and non-selective detectors there
is a clear advantage to the simplified KLM \textsc{cnot} gate, where
the inherent symmetries in the gate allow the use of two independent
SPDC sources to supply the control, target and ancillary photons, with
errors from the use of non-Fock states making little contribution.
The other two implementations suffer from errors introduced by the
non-Fock state inputs which cannot be postselected out. While the
situation may be mitigated somewhat by using a weak SPDC source this
would occur at the expense of the count rate of valid events that may
be collected from the gate. 

The conclusion we arrive at is that an experimental program focusing
on the simplified KLM \textsc{cnot} gate would then allow immediate
characterisation of the gate with current sources and detectors, with
the operation of the gate in a non-destructive fashion becoming
possible when single photon sources and selective detectors become
available.

We would like to acknowledge support from the the Australian Research
Council and the US Army Research Office. AG was supported by the New
Zealand Foundation for Research, Science and Technology under grant
UQSL0001. WJM acknowledges support for the EU project RAMBOQ. We would
also like to thank Michael Nielsen, Jennifer Dodd, Nathan Langford,
Tim Ralph and Gerard Milburn for helpful discussions.

\bibliographystyle{prsty}


\end{document}